# Higher harmonics of ac voltage response in narrow strips of $YBa_2Cu_3O_7$ thin films: Evidence for strong thermal fluctuations

JG Ossandon[a,c], S Sergeenkov[a,b], P Esquinazi[c] and H Kempa[c]

[a]*Department of Engineering Sciences, Universidad de Talca, Curicó, Chile*
[b]*Bogoliubov Laboratory of Theoretical Physics, Joint Institute for Nuclear Research, 141980 Dubna, Russia*
[c]*Abteilung Supraleitung und Magnetismus, Universität Leipzig, Linnéstrasse 5, D-04103 Leipzig, Germany*

*ABSTRACT*

*We report on measurements of higher harmonics of the ac voltage response in strips of $YBa_2Cu_3O_{7-\delta}$ thin films as a function of temperature, frequency and ac current amplitude. The third (fifth) harmonic of the local voltage is found to exhibit a negative (positive) peak at the superconducting transition temperature and their amplitudes are closely related to the slope (derivative) of the first (Ohmic) harmonic. The peaks practically do not depend on frequency and no even (second or fourth) harmonics are detected. The observed data can be interpreted in terms of ac current induced thermal modulation of the sample temperature added to strong thermally activated fluctuations in the transition region.*

**1. Introduction**

The measurement of higher harmonics created by ac transport currents or ac magnetic fields has been used as a tool to study different phenomena, both fundamental and related to applications, in superconducting materials. For example, the estimate of the order parameter relaxation time [1], probe of a chiral glass phase [2] and π–junction Josephson behavior [3] in granular superconductors, as well as detection of superconducting phases with different critical temperatures $T_c$ in inhomogeneous samples. Recently, special attention has been paid to the role of non-ohmic harmonics in thermodynamic and transport properties of high-temperature superconductors (HTS) [4-8] and in a contactless inductive method to measure critical currents in superconducting films [9-11], which complements the nonlinear ac susceptibility first harmonic method to measure critical currents [12]. Moreover, some interesting theoretical predictions have been made regarding third harmonics generation based on the anisotropic scattering rate of normal charge carriers [13,14] as well as produced by electric field modified fluctuation effects in the electrical conductivity [15] or in the thermal response of the film to an harmonic current [16].

In the present paper, we report on measurements of the first, third and fifth harmonics in narrow strips of $YBa_2Cu_3O_7$ (YBCO) HTS films and discuss their temperature and ac-current amplitude dependence. We provide a satisfactory theoretical interpretation of the observed phenomena assuming an ac current induced modulation of the sample temperature added to strong thermally activated fluctuations in the transition region that drive the sample in the normal state. This paper is divided in four more sections. In the next section we describe the experimental and sample details. In section 3 we present the main results and in section 4 the model used to describe the data is presented. We compare this model with the data in section 5 and a summary is written in section 6.

## 2. Experimental and sample details

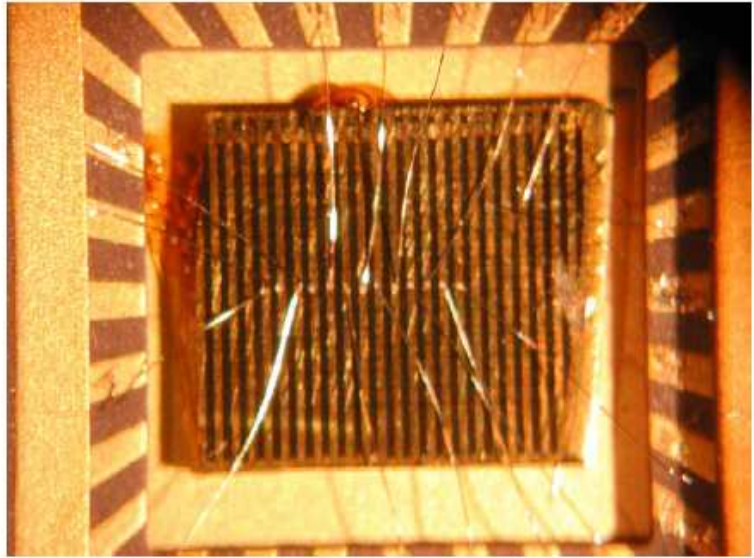

Figure 1. A top view of the typical sample I/3/1 LS.

To study the ac-response of a YBCO thin-film strip in the vicinity of the superconducting transition, the sample was mounted on the cold head of a standard, non optical, closed cycle refrigerator inside a vacuum bell. A dual channel auto-tuning temperature controller was used to control the temperature measured by a Pt-thermometer, mounted on the sample frame, and a silicon diode thermometer, mounted on the cold head. The temperature was controlled with a precision of ~ 50mK in the range of a few degrees above and below $T_c$. A mechanical vacuum pump provided the necessary heat insulation and the whole sample holder was shielded from the room temperature radiation by a metallic cylinder attached to the cold head.

A lock-in amplifier (Stanford Research System, model SR-830 DSP) was used both as a wave generator and as a nanovoltmeter. We restricted ourselves to low-frequency measurements, from 0.33 Hz to 66.6 Hz, to allow for thermal relaxation of the sample. In order to achieve a stationary ac-current source, free of inductive effects, the output ac signal of the lock-in amplifier was connected to a Voltage-Current Converter (Valhalla Scientific, model VS-2500), from which, alternatively, various fixed, constant amplitude, ac-transport currents, in the range from 0.5 mA to 20 mA (rms-values), were applied through the sample electrodes. A digital oscilloscope, working in dc-mode, was continuously monitoring the current amplitude. The ac-voltage response of the YBCO thin film sample was filtered and measured by the Lock-in amplifier.

The YBCO samples for this experiment were two rectangular pieces, 4x10 mm$^2$ size, both cut from a circular wafer (labeled S082). The plate consisted of a 470 μm thick, sapphire ($Al_2O_3$) substrate, of 76 mm diameter, having a ~240 nm thick YBCO film deposited by laser ablation on either side of it [17]. For better adhesion, a ~10 nm thick cerium oxide layer was interposed between the YBCO films and the substrate. A 200 nm thick gold layer covered both YBCO films on either side of the plate, which we further patterned for electric contacts. Using a diamond cutter, the two rectangular pieces were cut and then labeled I/3/1 and II/3/2/1 respectively, according to their original location on the S082 wafer.

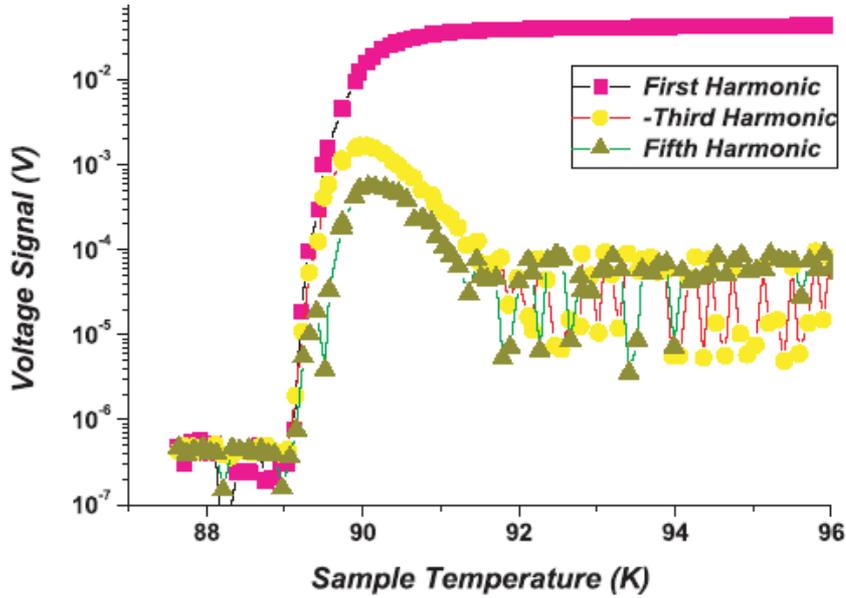

**Figure 2.** Temperature dependence of the first, third and fifth harmonics of the local voltage along strip 1 in sample II/3/2/1 LS, for an applied ac current of 10 mA (rms amplitude) and frequency of 3.33 Hz.

Using chemical etching and photolithographic techniques, the gold layer on each sample was partially removed and appropriately patterned to form a gold mask, consisting of a parallel array of golden fringes, each fringe being 100 microns wide and 100 microns apart from each other, similar to that used for non-local measurements [18], see Fig. 1. Finally, and in contrast to the samples measured in [18], in the other direction perpendicular to the golden fringes, a parallel array of deep wrinkles was cut in order to create long, narrow YBCO-film strips of constant width, having on their surface a uniform sequence of golden electrodes available for electric contact, see Fig.1.

The transverse-to-the-golden-fringes wrinkles array was carefully cut using two different techniques [19]: a laser beam method in sample II/3/2/1, and an ion-beam method in sample I/3/1. The separation between parallel, adjacent wrinkles was chosen to be, alternatively, 100, 150, and 200 microns, to test for eventual geometrical effects. The mean profile, or cross section, of each wrinkle was ~10 microns wide, and from 500 microns (ion beam method) to 600 microns deep (laser beam method); in both cases deep enough to cut through the three layers (gold mask, YBCO film and $CeO_2$ buffer layer) and penetrating all the way down to the insulating substrate, so that the YBCO strips were electrically isolated from each other. Atomic force microscopy (ATM) was used to check the depth profile of the wrinkles array. Moreover, direct continuity testing was conducted to check for eventual leaks from one strip to another.

Due to their large dimensions, as compared to the sample holder, the rectangular plates were separated in two halves, each piece having ~4x5 mm$^2$, and labeled respectively, I/3/1 RS, I/3/1 LS, II/3/2/1 RS, and II/3/2/1 LS. The piece was glued with varnish to the base of a chip carrier frame. Around the frame, 28 gold electrodes were available to make wiring connections to the surface contacts. A top view of sample I/3/1 LS is shown in Fig. 1. Using ultrasound bonding several thin aluminum wires of ~50 μm diameter were bonded to bridge some frame electrodes to appropriate locations on the golden mask. A 200 microns wide strip ("Strip 1") on sample II/3/2/1 LS was wired with eight point contacts in a row, for a total length of 2.1 mm. Depending on the choice of the transport current leads, this configuration allowed to conduct "local" (inside the current flow), as well as "non-local" (beyond the current flow) voltage measurements along the narrow strip to check for the excess voltage singularities near the transition temperature [18]. Similarly, another strip of 0.150 mm width and 1.9 mm length ("Strip 3") having seven point contacts in a row was prepared on the same sample. The discussion of the non-local results is beyond the scope of this paper. Nevertheless we would like to stress that the non-local signals were negligible due to the small width of the strips, which reduces significantly the inhomogeneous distribution of the input current in agreement with the interpretation given in [18].

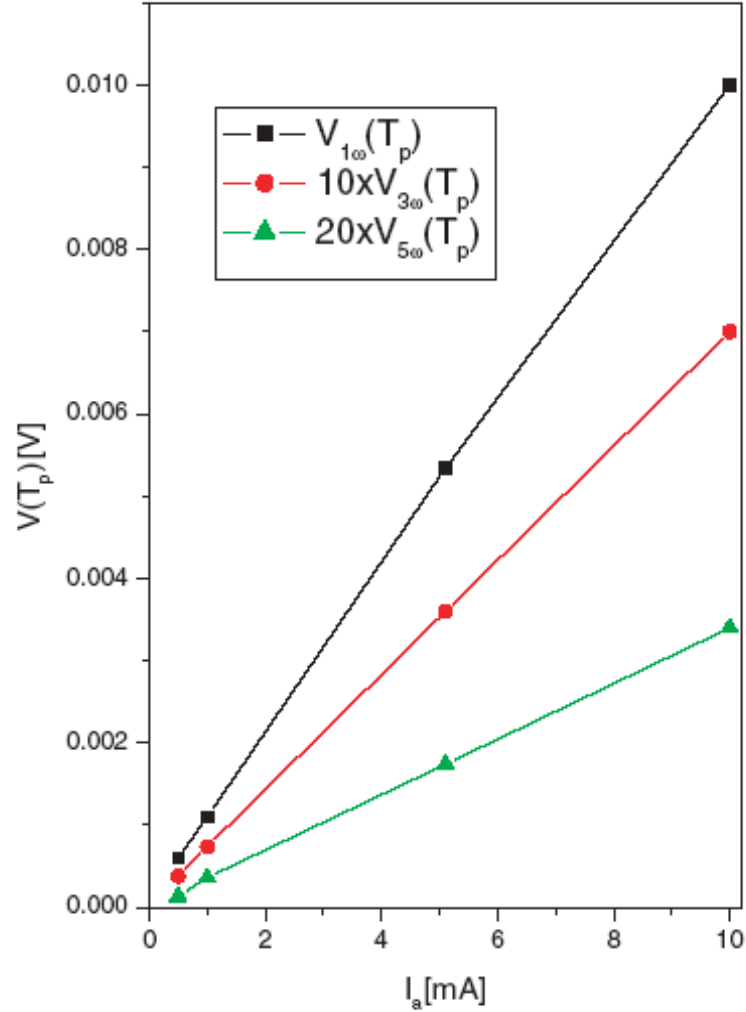

**Figure 3.** The maximum of the first, third and fifth harmonics in the transition interval versus the ac current amplitude, as measured on strip 3 of sample II/3/2/1 LS.

The experiment consisted in measuring the isothermal local voltage response of the YBCO thin film strips as a function of temperature in the vicinity of the superconducting transition. A four point configuration was chosen to introduce the current into the strip (using two outer electrodes) and to read the voltage (using two inner electrodes). The applied current was in each case a sinusoidal wave having alternative amplitudes of 0.5, 1, 5, 10 and 20 mA (rms-values) and frequencies of 3.33, 16.66, 33.3 and 66.6 Hz, while the sample temperature was set to be typically in the interval from 85 K to 95 K, stabilized within 0.05 K at each data point, with sweeping steps of 0.1 K. The frequency values were chosen far from any simple multiple or sub-multiple of the 50 Hz basic line frequency in order to avoid disturbing couplings or resonance effects.

## 3. Experimental results

First, third and fifth harmonic amplitudes of the voltage response were routinely measured by the lock-in amplifier as a function of temperature near the superconducting transition. The third (fifth) harmonic was clearly observed to have a negative (positive) peak right at the fluctuation interval. Only odd harmonics showed up; no even harmonics were detected. Figure 2 shows a typical case: it depicts on a logarithmic scale the first, third and fifth harmonics signals as a function of temperature of the local voltage along strip 1 in sample II/3/2/1 LS, for an applied ac-current of 10 mA (rms-amplitude) and frequency of 3.33 Hz. Due to the logarithmic scale, the negative sign of the third harmonic was reversed. In the transition interval, which extends from 92.0 K (onset-$T_c$) down to 89.0 K, the third and fifth harmonics signals show clear, positive peaks, whose amplitudes seem to be roughly proportional to the slope of the first harmonic signal. Also, using different applied currents, it was observed that the amplitudes of the first, third and fifth harmonics were closely proportional to the current amplitudes and independent of the chosen frequencies. Figure 3 shows the voltage maxima for all three harmonics vs the ac-current amplitude. As it is clearly seen, all the data follow a *linear* dependence.

It is worth mentioning that recently Chéenne et al. [20] found similar results for the third harmonic, using similar narrow YBCO strips, experimental set-up and measuring technique. However, in contrast with our observations, their data do not exhibit any universal behavior as far as the current dependence of the harmonic maximum is concerned.

**4. Interpretation**

It is already well established (see e.g. [15,16,20] and further references therein) that an ac current $I(t) = I_a \sin \omega t$ applied to a superconducting film not only produces an ac voltage response $V = R(T)I$ (where $R(T)$ is the temperature dependent ac resistance of the film) but simultaneously leads to temporal oscillations of the sample temperature $T(t) = T_0 + \delta T(t)$ around its steady value $T_0$. The dissipation power in the film can be expressed by $P(t) = G\delta T(t)$, where $G$ is the thermal boundary conductance between the strip and the substrate. To proceed with the calculation of the harmonic response of a superconducting strip we need an explicit expression for the temporal variation of the temperature, $\delta T(t)$. It is natural to assume that the resulting Joule heating of the electron system is given by the dissipation power $P(t) = VI = R(T)I^2$. Expanding the resulting resistance around the equilibrium temperature $T_0$ (assuming that $T_0 >> \delta T$), we have $R(T) = R_0 + R'_0 \delta T + \frac{1}{2} R''_0 \delta T^2$ with $R_0 = R(T_0)$, and $R'_0 = R'(T_0)$ and $R''_0 = R''(T_0)$ being the first and second temperature derivatives of the resistance. As a result, we obtain $\delta T(t) \cong \Theta \sin^2 \omega t$ for the time dependence of the temperature variation, where

$$\Theta = \frac{1}{2\pi} \int_0^{2\pi} d(\omega t) \delta T(t) = \frac{I_a^2 R_n}{G}, \quad (1)$$

is a characteristic temperature amplitude related to ac current induced phenomena and $R_n$ is the value of the resistance $R_0$ in the critical state, i.e. at $I_a = I_c(T)$, see below.

Making use of the well-known trigonometric relations

$$\sin^3 x = \frac{3}{4} \sin x - \frac{1}{4} \sin 3x, \quad (2)$$

and

$$\sin^5 x = \frac{1}{16} \sin 5x - \frac{5}{16} \sin 3x + \frac{10}{16} \sin x, \qu(3)$$

we get for the ac voltage response of the film:

$$V(t) = V_{1\omega} \sin \omega t + V_{3\omega} \sin 3\omega t + V_{5\omega} \sin 5\omega t \tag{4}$$

where the first three odd harmonics are:

$$V_{1\omega} = R_0 I_a \left( 1 + \frac{3R'_0}{4R_0} \Theta + \frac{5R'^2_0}{8R_0{}^2} \Theta^2 + \frac{5R''_0}{16R_0} \Theta^2 \right), \tag{5}$$

$$V_{3\omega} = -\frac{\Theta R'_0}{4} I_a \left( 1 + \frac{5R'_0}{4R_0} \Theta + \frac{5R''_0}{8R'_0} \Theta \right), \tag{6}$$

$$V_{5\omega} = \frac{\Theta^2 R'^2_0}{16R_0} I_a \left( 1 + \frac{R_0 R''_0}{2R'^2_0} \right). \tag{7}$$

Notice that the third harmonic is negative and all harmonics do not depend on frequency, in agreement with our observations. Moreover they depend only *linearly* on the amplitude of the ac current $I_a$. To test the dependence on the characteristic temperature amplitude $\Theta$ (see Eqs.(5)-(7)) one needs to perform accurate thermopower measurements since the dissipation power $<P(t)>=G\Theta$ produced in the film could give rise to observable thermoelectric effects with an average thermal voltage $V_{th}=S\Theta$ (where $S$ is a proper Seebeck coefficient).

In the presence of strong thermal fluctuations around $T_c$, it is quite reasonable to approximate the resistance by a thermally activated expression of the form [21,22]

$$R_0(T, I_c) = R_n I_0^{-2}[\gamma(T, I_a)] \tag{8}$$

where $I_0(x)$ is the modified Bessel function of zero order, and

$$\gamma(T, I_a) = \left(1 - \frac{\Phi}{\Phi_0}\right)^2 + \frac{U(T, I_a)}{2k_B T} \tag{9}$$

is the normalized barrier height with

$$\Phi = \Phi_0 + L[I_a - I_c(T)] \tag{10}$$

being the total flux through the film (in the vicinity of the critical-state regime), and

$$U(T, I_a) = \frac{1}{2} L[I_a - I_c(T)]^2 \tag{11}$$

the current-induced activation energy; $L$ is the inductance of the film, and $\Phi_0$ is the flux quantum. The first term in the rhs of Eq.(9) stands for a quantization (Little-Parks type) condition in a strip [23].

For the temperature dependence of the critical current in the fluctuation regime we assume a simple London relation [23] of the form

$$I_c(T) = I_c(0)\left(1 - \frac{T^2}{T_c^2}\right) \quad (12)$$

This is plausible since for the critical current density in a narrow thin film of thickness $d$ it is reasonable to assume [12,23] $j_c = \pi h_c/d$, with $h_c(T) = h_c(0)(1-t^2)$ and $t = T/T_c$.

As a result, the normalized activation energy (which enters the fitting expression (8)) can then be rewritten as follows:

$$\gamma(T, I_a) = \gamma_0 \left(\alpha - \frac{T^2}{T_c^2}\right)^2 \quad (13)$$

where

$$\gamma_0 = \frac{L^2 I_c^2(0)}{\Phi_0^2}\left(1 + \frac{\Phi_0^2}{2k_B T_C L}\right) \quad (14)$$

and

$$\alpha = 1 - \frac{I_a}{I_c(0)} \quad (15)$$

## 5. Comparison with the model and numerical results

Fig. 4 shows the fitting of the typical data (normalized voltage harmonics $V_{n\omega}/V_{n\omega,\max}$ as a function of reduced temperature $T/T_c$) for $I_a = 1mA$ according to Eqs. (5)-(15). (We notice that as usual [20] the raw data were smoothed using the standard Savitzky-Golay filtering procedure). The agreement between the theoretical model and the actual data is remarkably good in all cases. The sign reversal of the third harmonic signal is also taken into account. The fitting requires only two model parameters to be adjusted: $\gamma_o$ and $\alpha$. The theoretical curves are highly sensitive to the value of $\alpha$. Even small relative variations in $\alpha$ (of the order of 0.1 %) readily affect the optimal fitting values.

The best fit through the data points produce the following values for the model parameters: $\gamma_o$ =775 (for all currents and all harmonics, with ~1% accuracy), $\alpha$ (1mA)=0.999, $\alpha$ (5mA)=0.995, $\alpha$ (10mA)=0.990 (also for all three harmonics, with 0.1 % accuracy). In excellent agreement with Eq.(15), the $\alpha$ values follow linear dependence on applied current $I_a$. Using these fitting parameters, we can estimate the typical values of zero temperature critical current $I_c(0)$ and film inductance $L$. The results are as follows: $I_c(0) \approx 1.0A$ and $L \approx 0.32pH$, respectively. Notice that the estimate of the film inductance is very close to its geometrical value $L=\mu_0 d=0.3pH$ assuming that the main events affecting the behavior of our film are related to its thickness $d=240nm$ (here $\mu_0$ is magnetic permeability).

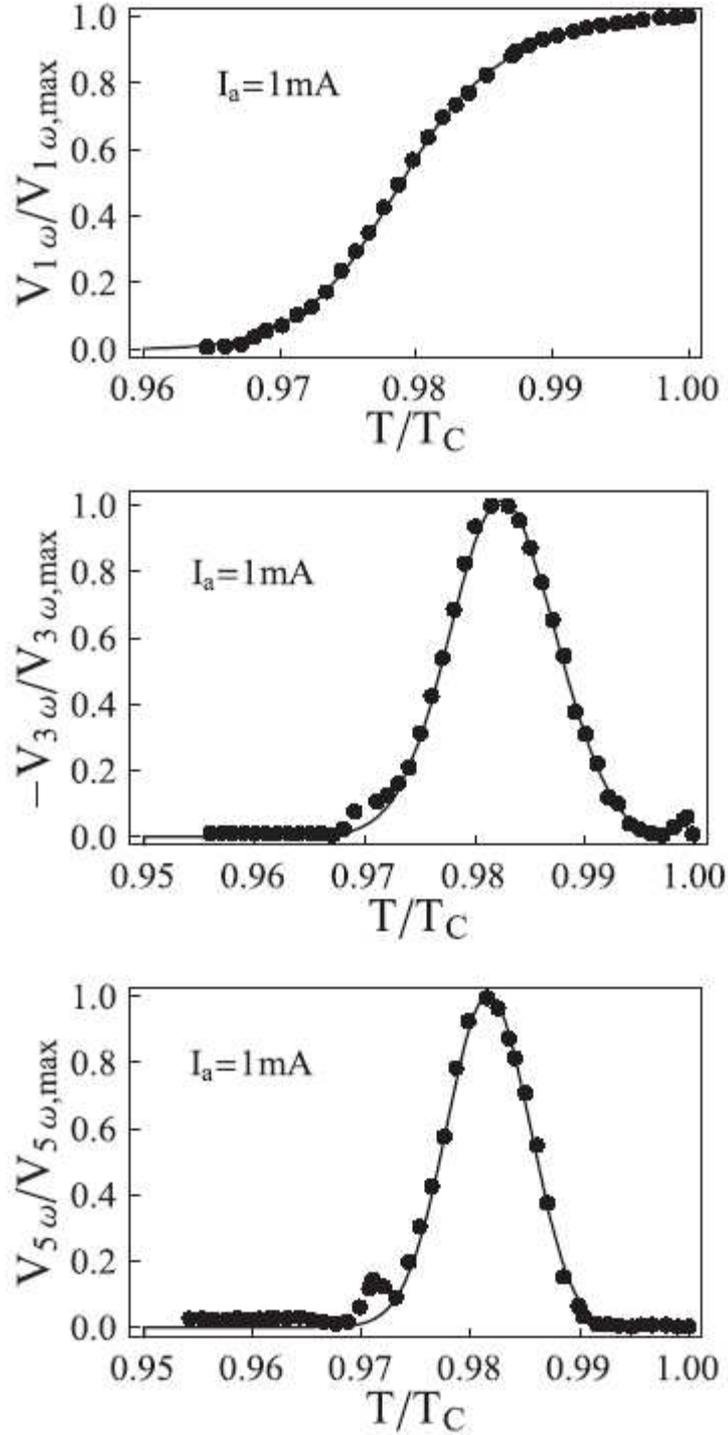

**Figure 4.** The fitting of the typical data for $I_a = 1$ mA ac current amplitude according to equations (5)–(15).

Besides, using the experimental values of $V_{1\omega}$ (90K), $V_{3\omega}$ (90K) and $V_{5\omega}$ (90K) for $I_a = 10mA$, from Eqs.(5), (6), and (8) we obtain $R_n \approx 1.2 Ohm$ and $\Theta \approx 1mK$ for the estimates of the critical state resistance and the characteristic temperature, respectively. As is expected, this temperature may produce quite a tangible thermoelectric effect in the strips, equivalent to a thermal voltage of the order of $V_{th}=S\Theta \approx 1nV$ (given the typical value of the Seebeck coefficient $S \approx 1\mu V/K$ in YBCO films). Finally, from Eq.(1) we estimate the value of the thermal boundary conductance between the strip and the substrate to be $G \approx 0.1 W/K$. Thus, the corresponding thermal current $I_G = (G/R_0')^{1/2} \approx 0.5A$ (with $R_0'(T_c) \approx 0.4\ Ohm/K$ according to Eq.(8)) is much larger than the maximum ac currents used in our experiments ($I_a = 20mA$). This, in turn, guarantees that our films were far from the so-called [20] thermal runaway conditions during the actual measurements.

## 6. Conclusion

The experimental results for measurements of higher harmonics of ac voltage response in the strips of $YBa_2Cu_3O_{7-\delta}$ based thin films as a function of temperature, frequency and ac current amplitude were presented and interpreted in terms of an ac current induced thermal modulation of the sample temperature added to strong thermally activated fluctuations in the transition region. The fit of the model to the experimental data yielded precise information of some film properties such as critical current, film inductance, critical state resistance and thermal boundary conductance.

*ACKNOWLEDGMENTS*

The authors are indebted to Herr Michael Lorenz (who made the films), Dr. Michael Ziese for helpful collaboration in the experiments, and Ms. Eva Salamatin (IOM, Leipzig) for technical assistance. We thank David Christen (Oak Ridge National Laboratory) for careful reading of the manuscript and very useful comments. The work of J.G.O. was possible thanks to a sabbatical permit from the Universidad de Talca and was partially supported by the Deutscher Akademischer Austauschdienst (DAAD) and the Chilean Fondo Nacional de Ciencia y Tecnología (FONDECYT, grant # 1040666). Participation of S.S. was made possible through an invitation of the Universidad de Talca.